# Benchtop Nonresonant X-ray Emission Spectroscopy: Coming Soon to Laboratories and XAS Beamlines Near You?

**Devon R. Mortensen[1], Gerald T. Seidler*[1], Alexander S. Ditter[1] and Pieter Glatzel[2]**

[1]Deparment of Physics, University of Washington, Seattle, WA 98195, USA
[2]European Synchrtron Radiation Facility, CS40220, Grenoble 38043 Cedex 9, France

*seidler@uw.edu

**Abstract**. Recently developed instrumentation at the University of Washington has allowed for nonresonant x-ray emission spectra (XES) to be measured in a laboratory-setting with an inexpensive, easily operated system. We present a critical evaluation of this equipment by means of Kβ and valence-level XES measurements for several Co compounds. We find peak count rates of ~5000/s for concentrated samples and a robust relative energy scale with reproducibility of 25 meV or better. We furthermore find excellent agreement with synchrotron measurements with only modest loss in energy resolution. Instruments such as ours, based on only conventional sources that are widely sold for elemental analysis by x-ray fluorescence, can fill an important role to diversify the research applications of XES both by their presence in non-synchrotron laboratories and by their use in conjunction with XAFS beamlines where the complementarity of XAFS and XES holds high scientific potential.

## 1. Introduction

Nonresonant x-ray emission spectroscopy (XES), while far less common as ubiquitous as x-ray absorption spectroscopy (XAS), has seen steady growth in a number of fields over the past several decades. In particular, Kβ (3p→1s) and valence-level XES has seen extensive use in determining spin state, oxidation state, and ligand speciation in 3d transition-metal materials.[1-3] Because this technique probes the emissions resulting from the filling of coreholes, it provides complementary information to XAS and combination of the two spectroscopies is a natural and powerful extension.

Unfortunately the infrastructure for XES has lagged considerably behind that for XAS with only a handful of synchrotron end stations dedicated to such measurements, see for example [4-7]. This access scarcity has significantly hampered the further development of XES as a routine complement to XAS. We have recently demonstrated, however, that benchtop instrumentation based on high-resolution spherically-bent crystal analyzers (SBCAs) can in many cases provide synchrotron-quality XES measurements.[8] Here we explore the capability and limitations of this laboratory-based equipment via a direct comparison to synchrotron data and we also report on an instrumental improvement that helps to finely anchor the accuracy of the energy scale.

## 2. Experimental Details

Laboratory spectra were recorded using a Rowland-circle monochromator recently developed at the University of Washington [8], see figure 1. X rays are generated by a conventional tube source (Moxtek) with an Au anode operated at 40 kV bias and 200 μA current, i.e., only 8 W total electron beam power.

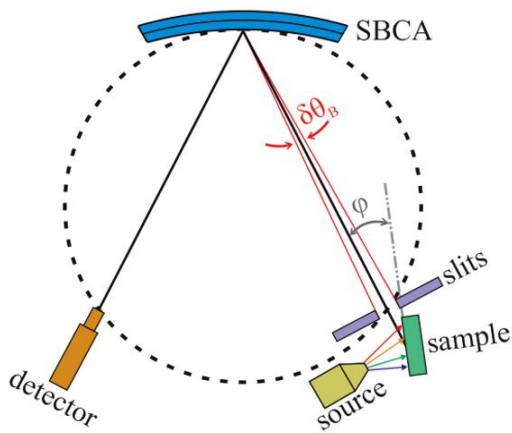 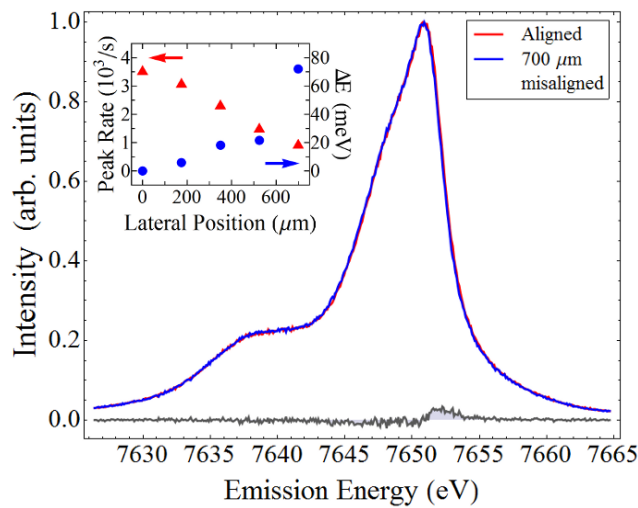

**Figure 1.** The lab-based Rowland-circle spectrometer.[4] The slits in front of the sample constrain the angular width ($\delta\theta_B$) probed by the analyzer which improves energy resolution and reproducibility. The sample is slightly rotated ($\phi$) to allow improved line-of-sight to the SBCA.

**Figure 2.** Peak-intensity normalized CoO K$\beta$ spectra measured at the aligned and the extreme misaligned location. The gray curve shows the residual between the two spectra. Inset: the peak count rate and energy shift as a function of lateral sample position.

As has been emphasized elsewhere,[8] the small anode-to-sample distance possible with this type of source allows the sample to subtend a very large solid angle, giving highly efficient excitation that can reach rates intermediate between those typical for a monochromatized BM and ID at a third-generation synchrotron. The sample fluorescence was analyzed and refocused at the detector position by a 1-m radius of curvature Ge (444) SBCA (XRS Tech). A geometric correction is applied to compensate for the Bragg angle dependence of the height of the refocused spot at the finite-height (5-mm) detector.

The sample is rotated slightly (~15°) to provide improved line-of-sight to the analyzer for the fluorescence while still allowing for minimal offset from the x-ray tube anode (~3 mm). Most recently, instead of placing the sample directly on the Rowland geometry we use a narrow entrance slit on the Rowland circle and move the sample + source subassembly ~10mm behind the slit. The purpose of this slit is two-fold. First, the dominant limit on energy resolution is the apparent angular width of the illuminated sample area as seen by the SBCA. The entrance slit gives some ability to tune the resolution by adjusting its width. Higher resolution of course comes at the cost of lower count rates and the optimal tradeoff is determined on a case-by-case basis. For this study a slit width of 0.5 mm was selected. This corresponds to an angular width of 0.03° (0.48 eV) at 82.93° Bragg angle (7650 eV).

The second purpose of the entrance slit is more subtle and involves the difficulty of reproducing exact sample position. Since the relative energy scale is determined geometrically, a small lateral misalignment of the sample will register as an incorrect energy shift in the resulting spectrum. This ambiguity presents a significant barrier to cross-comparisons between different samples, which is especially pressing in the case of K$\beta$ XES from 3d transition metal compounds where shifts in the K$\beta_{1,3}$ peak can indicate changes in spin state and other local electronic properties.[1] With the slit constraining the angular range being sampled by the SBCA, however, this becomes a minor issue; a misalignment large enough to meaningfully shift the energy scale will fall outside of the sampled range, resulting in an obviously reduced signal. For means of example, we measured CoO K$\beta_{1,3}$ at several lateral sample positions as well as a spectrum where the sample was removed, replaced, and realigned – see figure 2. These results demonstrate a highly robust energy scale that is insensitive to exact sample position. The inset of figure 2 shows that shifts in the energy scale as small as ~25 meV are accompanied by factor of 2 decreases in overall intensity. Simple optimization of signal intensity with respect to the sample position behind the entrance slit therefore ensures outstanding consistency of energy scale between different measurements.

For the study below, synchrotron measurements were performed at beamline ID26 of the European Synchrotron Radiation Facility (ESRF). The incident energy was tuned to 7.8 keV using a double Si (111) monochromator. The total incident flux was ~ 2 x 10$^{13}$ photons/s. Samples were oriented at 45° in the scattering plane, resulting in a beamsize of 0.5 mm in-plane and 1 mm out-of-plane. The resulting fluorescence was analyzed by an array of five Ge (444) SBCAs in a 1-m Rowland circle geometry. The energy resolution, which varies slightly between SBCAs, is ~1.1 eV as determined from the FWHM of elastic scatter lines.

Samples were prepared from high purity powder (99.99% $Co_3O_4$; 99.5% $LiCoO_2$) from Alfa Aesar. The same samples were measured at both the University of Washington lab and the ESRF, thus eliminating possible systematic effects due to variations in sample preparation.

## 3. Results and Discussion

In figure 3 we present a laboratory/synchrotron comparison of Kβ XES for $Co_3O_4$ and $LiCoO_2$. To better overlay with the laboratory data the ESRF spectra have been additionally broadened by convolution with a Gaussian of FWHM = 0.8 eV. This is larger broadening than expected from the finite size of the entrance slit discussed above and is mostly likely due to the SBCA being slightly off circle in the laboratory system. Nonetheless the net effective energy resolution of ~1.4 eV is more than sufficient to resolve the key XES features without loss in scientific merit.

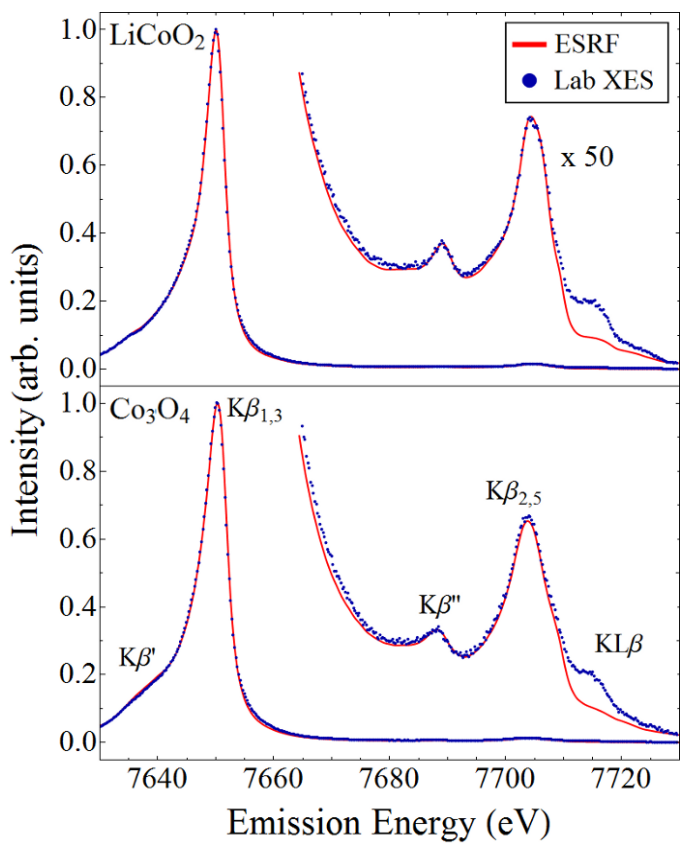

**Figure 3.** Kβ XES for $LiCoO_2$ and $Co_3O_4$ powder samples. The red curve shows spectra collected as ESRF; blue dots show data taken using the laboratory-based instrument described in the text. All spectra have been normalized to peak intensity. For ease of presentation, the valence-to-core emission is shown amplified 50x.

The two datasets are in excellent agreement for both the main Kβ (Kβ$_{1,3}$ and Kβ') and valence-to-core (Kβ$_{2,5}$ and Kβ'') emission lines. The only discrepancy is the KLβ feature present only in the laboratory data. This emission is the result of a multi-electron excitation [9] and therefore exhibits a threshold behavior at roughly the K + L-edge binding energies. The excitation energy for the ESRF data (7.8 keV) is well below this threshold. The broadband tube source in the laboratory setup, however, emits x rays with energies up to 40 keV resulting in sizeable KLβ emission.

Despite the low-powered x-ray tube used here, the peak count rates at K$\beta_{1,3}$ are ~5 x $10^3$/s, allowing high quality spectra in very useful measurement times. Count-rate improvement of up to 100x could be realized by upgrading the x-ray source to a kW-level commercial tube of the type used commonly in fluorescence studies for elemental analysis. For comparison the ESRF measurements averaged ~2 x $10^6$/s/SBCA at peak.

## 4. Conclusion
We have demonstrated the effectiveness of inexpensive, easily assembled and operated benchtop equipment using only low-powered, conventional x-ray sources for high-resolution XES measurements. We find easy measurement of even the lowest-intensity valence-level emission lines for concentrated samples and also find excellent agreement with synchrotron-based studies. This versatile instrument design could have significant application in many fields where transition-metal and lanthanide compounds are of central importance, including electrical energy storage, f-electron chemistry, catalysis, and environmental chemistry. In addition to extending routine XES beyond the confines of synchrotron light sources, we propose that this type of user-friendly spectrometer could be used as an on-site supplement to XAFS beamlines allowing XES measurements without using synchrotron beamtime.

## 5. Acknowledgements
This work has been supported by: (1) the U.S. Department of Energy, Basic Energy Sciences under Grant No. DE-FG02-09ER16106 and also by the Office of Science, Fusion Energy Sciences and the National Nuclear Security Administration thought Grant No. DE-SC0008580; (2) the State of Washington through the University of Washington Clean Energy Institute; (3) The Joint Center for Energy Storage Research (JCESR), an Energy Innovation Hub funded by the U.S. Department of Energy, Office of Science, Basic Energy Science; and (4) The U.S. Department of Defense through a subcontract from Los Alamos National Laboratories.